\documentclass[useAMS,usenatbib]{mn2e}
\usepackage{graphicx}
\usepackage{amsmath}

\title[The $\lambda$~Bootis phenomenon]
      {The $\lambda$~Bootis phenomenon: interaction between a star and a diffuse interstellar cloud}
\author[Inga Kamp and Ernst Paunzen]{Inga Kamp$^{1}$\thanks{E-mail:
kamp@strw.leidenuniv.nl} and Ernst Paunzen$^{2}$\\
$^{1}$Leiden Observatory, PO Box 9513, NL-2300 RA Leiden, The Netherlands\\
$^{2}$Institut f\"{u}r Astronomie der Universit\"{a}t Wien,
          T\"{u}rkenschanzstr. 17, A-1180 Wien, Austria}
\begin{document}

\date{Accepted 2002 July 22. Received 2002 June 5}

\pagerange{\pageref{firstpage}--\pageref{lastpage}} \pubyear{2002}

\maketitle

\label{firstpage}

\begin{abstract}
          The $\lambda$~Bootis stars, a group of late B to early
          F-type population\,{\sc I} stars, have surface abundances that
          resemble the general metal depletion pattern found in the
          interstellar medium. Inspired by the recent result
          that the fundamental parameters of these peculiar stars differ 
          in no respect from a comparison sample of normal stars, the
          hypothesis of an interaction between a star and a diffuse
          interstellar cloud is considered as a possible explanation
          of the peculiar abundance pattern. It is found that such
          a scenario is able to explain the selective accretion of
          interstellar gas depleted in condensable elements as well
          as the spectral range of the $\lambda$~Bootis phenomenon.
\end{abstract}


\section{Introduction}

The $\lambda$~Bootis stars are late B to early F-type population\,{\sc I}
stars, which show a peculiar surface abundance pattern: while the light
elements (C, N, O and S) are roughly solar, the Fe-peak elements show
underabundances of up to a factor of 100. \citet{Venn}
are the first who noticed the similarity between this abundance pattern
and the depletion pattern of the interstellar medium (ISM) and
suggested the accretion of interstellar or circumstellar gas to explain 
the $\lambda$~Bootis stars. In this respect they differ from the
rest of the peculiar A-type stars where the abundance pattern is
caused by separation processes in the stellar atmosphere itself. The
abundances are ascribed to diffusion in the presence of 
slow rotation (Am stars) or strong magnetic fields (Ap stars).

\citet{Paunzen02} scrutinized the available observational 
data to put constraints on any model trying to explain the $\lambda$~Bootis 
phenomenon. A comparsion between the $\lambda$~Bootis stars and a reference 
sample of normal stars showed that both groups share the same 
fundamental parameters, like effective temperature, gravity, mass, rotational 
velocity and age. But most surprisingly, the Na abundance of 
the $\lambda$~Bootis stars revealed a correlation with nearby local interstellar
column densities of Na\,{\sc i}.

This discovery, although so far tentative, because of the inhomogeneity
of the stellar Na abundances, motivated a detailed analysis of
the interaction between a star and a diffuse ISM cloud as an
explanation for the $\lambda$~Bootis phenomenon.

\section{Interaction between ISM and star}
\label{model}

\citet{Vergely} reconstructed the density distribution of 
the ISM in the solar neighbourhood from Na\,{\sc i} and H\,{\sc i}
observations. They found density fluctuations of a factor of 1000
for Na\,{\sc i} and about 250 for H\,{\sc i}. Typical number densities 
for diffuse clouds in the ISM range from 0.1 to 100~cm$^{-3}$. 
 
\begin{figure}
\resizebox{\hsize}{!}{\includegraphics{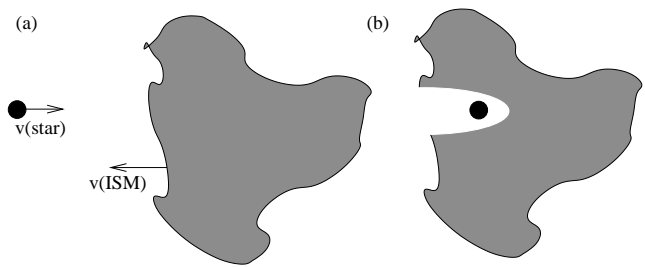}}
\caption{Sketch of a star encountering a diffuse ISM cloud (a)
and clearing a cavity inside (b)}
\label{fig_model}
\end{figure}

Fig.~\ref{fig_model} shows what is supposed to happen, 
if a $\lambda$~Bootis star travels
through a diffuse ISM cloud: the star creates a cavity in the ISM cloud.
There are two questions arising: What happens to the ISM dust? What are
the typical gas accretion rates? We assume in the following a moderate 
density of $n = 10$~cm$^{-3}$; using a mean molecular weight $\mu = 1.4$, 
appropriate for atomic gas, this leads to a lower limit of the ISM mass 
density, $\rho = 2.34\times 10^{-23}$~g~cm$^{-3}$.

The dust grains of the diffuse ISM cloud will be charged in
the vicinity of the star. Hydrogen will be mostly neutral
and the dominant charged gas species is C$^+$. Assuming 
50~\AA\  dust particles and a density of 
$n({\rm C}^+) = 10^{-3}$~cm$^{-3}$, we 
derive the distance at which coulomb forces and radiation 
pressure are equal
\begin{equation}
d = 9.47\times 10^3 \left(\frac{L_\ast}{L_\odot} \right)^{1/2}
                    \left(\frac{U}{\rm V} \right)^{-1/2}~~~{\rm AU}\,\,\,.
\end{equation}
The grain potential $U$ is related to strength of the stellar and
interstellar ultraviolet radiation field, the potential for dust 
grains at 500~AU around an A0\,V star being $U\sim 4-8$~V and
$U\sim 2-4$~V for a F5\,V star \citep{Weingartner} depending on the grain material. 
From this comparison we conclude that radiation pressure clearly
dominates at distances out to about 10\,000~AU from the star. 
At these low densities dust and gas are also not effectively 
coupled via collisions. Hence, we deal with dust and gas separately 
in the two following sections.

\subsection{Interstellar dust}

The ISM consists mainly of particles much smaller than 1~$\mu$m. The
following calculations are performed for 0.01~$\mu$m grains. \citet{Artymowicz} 
showed that the ISM dust particles undergo 
Rutherford scattering off the star and hence never approach the star closer 
than its avoidance radius
\begin{equation}
r_{\rm av} = \frac{2(\beta -1)GM_\ast}{v_\infty^2}~~~{\rm cm}\,\,\,,
\end{equation}
where $\beta$ is the ratio of radiation pressure to gravity.
For a spectral type of A5\,V (2\,R$_\odot$, 2\,M$_\odot$) which resembles
$\beta$~Pictoris and 0.01~$\mu$m 
grains, their fig.~4 gives an avoidance radius of approximately 500~AU. 
Therefore, the dust grains will not be accreted by the star.

\begin{table}
\caption{Stellar parameters used for the calculation of the avoidance radii.}
\begin{tabular}{lcccc}
\hline\\[-3mm]
\hspace*{5mm}Star  & A2\,V & $\beta$~Pic (A5\,V) & F2\,V & F5\,V \\[2mm]
\hline\\[-3mm]
$T_{\rm eff}$ [K]  & 9000  & 8000        & 6750  & 6550  \\
$M$ [M$_\odot$]    &  1.9  & 1.75        &  1.4  &  1.2  \\
$L$ [L$_\odot$]    & 14.1  & 8.7         &  3.5  &  1.7  \\
$L/M$              &  7.4  & 5           &  2.5  &  1.4  \\
$v_\infty$         & 10.7  & 8.6         &  6.7  &  5.9  \\
\end{tabular}
\label{tab_stellar_par}
\end{table}

Following their approach, we derived $r_{\rm av}$ for other stars
using the radiation pressure efficiencies (mixture of 50 per cent astronomical 
silicates and 50 per cent graphite) tabulated for various black body radiation 
fields by \citet{Draine}. Our result for $\beta$~Pictoris, $r_{\rm av}=340$~AU,
using an 8000~K black body radiation field,  and that of \citet{Artymowicz}, 
who used a \citet{Kurucz} model atmosphere, agree within a factor of 2, which is
more than satisfactory.

\begin{figure}
\resizebox{\hsize}{!}{\includegraphics{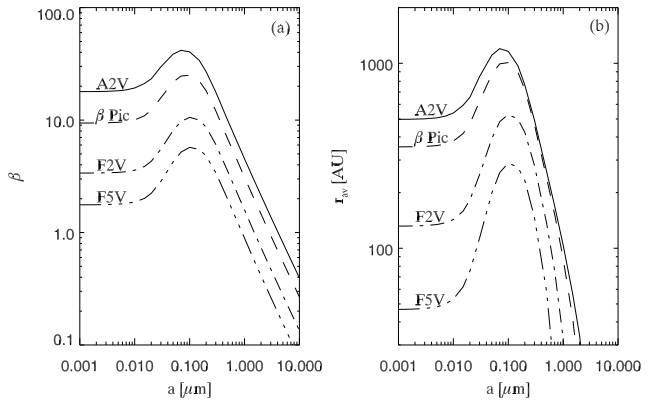}}
\caption{(a) Ratio of radiation pressure to gravity $\beta$ as a function of
dust grain size $a$ for a mixture of 50 per cent astronomical silicates and
50 per cent graphite. $\beta$ is shown for 4 different stars: A2\,V star (solid 
line) $\beta$~Pic (dashed line), F2\,V star (dash-dotted line) and F5\,V star 
(dash-dot-dotted line). (b) Radii of avoidance for different ISM dust grain 
sizes for the 4 different stars.}
\label{fig_beta_avoidance}
\end{figure}

Figure~\ref{fig_beta_avoidance} shows the ratio of radiation pressure to 
gravity for 4 stars: an A2\,V star, $\beta$~Pictoris, an F2\,V and an F5\,V star. 
The assumed stellar parameters are taken from \citet{Cox} and are 
shown together with the derived quantities in Table~\ref{tab_stellar_par}. 
The results clearly show that the avoidance radius for 0.01~$\mu$m grains 
around a main sequence F2 star is still about 130~AU, while it is only 30~AU
for a F5 star. Stars much cooler than that may not be able to prevent
accretion of ISM dust grains onto their surface.

\subsection{Interstellar gas}

Interstellar gas around O and B-type stars is known to be ionized (H\,{\sc ii} 
regions). In a first approximation, the Str\"{o}mgren radius around the 
star can be calculated from the number of Lyman~$\alpha$ continuum photons, 
$N_{\rm u}$ and the density of the ISM, $n_{\rm H}$,
\begin{equation}
r_{\rm s} = 8.66\times 10^{-13} \left(\frac{R_\ast}{R_\odot} \right)^{2/3}
            N_{\rm u}^{1/3} n_{\rm H}^{-2/3}~~~{\rm AU}\,\,\,.
\end{equation}
While the radiation field of an O star can be reasonably approximated
by a black body, A-type stars have strong line blanketing in the UV. Hence, 
to calculate the number of Lyman~$\alpha$ continuum photons for A and 
F-type stars, we used \citet{Kurucz} model atmospheres and the 
appropriate stellar radii taken from \citet{Cox}
\begin{equation}
N_{\rm u} = \int_0^{912 \rm \AA} F_\nu d\nu
                 ~~~{\rm photons~cm^{-2}~s^{-1}}\,\,\,.
\end{equation}
The resulting Str\"{o}mgren radii are summarized in Table~\ref{tab_r_s}
for various A and F-type stars. Assuming a density of 10~cm$^{-3}$ for
the ISM cloud, the Str\"{o}mgren radii lie between 3.1 and 0.8~AU.
Assuming a relative velocity of 17~km~s$^{-1}$, a gas particle takes
about 1~yr to reach the stellar surface from this radius. 
The time which a neutral hydrogen atom spends before it is ionized 
can be calculated from the photoionization rate of H\,{\sc i}
\begin{equation}
t = R^{-1} = \left(\int_{\rm 912 \AA}^\infty a_{\rm E} F_\nu \left( \frac{r}{R_\ast} \right) ^2 
                                       d\nu \right)^{-1}~~~{\rm s}\,\,\,,
\end{equation}
using the parameterized photoionisation rate $a_{\rm E}$ from Cox (2000).
The timescales at the Str\"{o}mgren radius are then between 
$3.2\times 10^8$ and $10^9$~yr depending on the spectral type of the star. 
A comparison with the dynamical timescale shows that the accreted 
interstellar gas will be mostly neutral.

\begin{table}
\caption{Str\"{o}mgren radii for various A and F-type stars, assuming
         a recombination rate appropriate for H\,{\sc ii} regions
         ($T=10^4$~K). The photoionisation rate $R$ is calculated at the
         Str\"{o}mgren radius $r_{\rm s}$.}
\begin{tabular}{lllrr}
\hline\\[-3mm]
Star & $T_{\rm eff}$ [K] & $R_\ast$ [R$_\odot$] & 
$r_{\rm s}$ [$n_{\rm H}^{2/3}$~AU] & $R$ [10$^{-17}$~s$^{-1}$]\\[2mm]
\hline\\[-3mm]
A2 & 9000              & 1.56                 &103.83 \hspace*{3mm}  & 77.77\hspace*{5mm}\\
A4 & 8500              & 1.51                 & 32.32 \hspace*{3mm}  & 23.92\hspace*{5mm}\\
A6 & 8000              & 1.48                 & 13.23 \hspace*{3mm}  &  9.84\hspace*{5mm}\\
A8 & 7750              & 1.45                 &  8.54 \hspace*{3mm}  &  6.37\hspace*{5mm}\\
A9 & 7500              & 1.43                 &  5.45 \hspace*{3mm}  &  4.07\hspace*{5mm}\\
F0 & 7250              & 1.40                 &  3.77 \hspace*{3mm}  &  2.83\hspace*{5mm}\\
\end{tabular}
\label{tab_r_s}
\end{table}

\citet{Bondi} have calculated the mass accretion rate of
material onto a star as it moves through an ISM cloud. The basic
assumptions are that (1) the gas is mainly molecular and hence the
temperature remains small via effective cooling and (2) effects
of gas pressure are small compared to the gravitational forces.
The first assumption may pose a problem for an A-type star which
efficiently dissociates the molecular ISM gas in its vicinity, but
C, C$^+$ and O are efficient alternative coolants of the gas
\citep{Black}.
The second point can be checked by calculating the critical radius 
up to where gas pressure effects can be neglected \citep{Bondi}
\begin{equation}
r_{\rm crit} = \frac{GM}{u^2}~~~{\rm cm}\,\,\,.
\end{equation}
Here $u$ is the thermal velocity of the interstellar material. Assuming
$u = 0.1$~km~s$^{-1}$ leads to a critical radius of $8.9\times 10^4$~AU
for a solar-mass star.
This is well outside the regime we consider in this study.

\begin{figure}
\resizebox{\hsize}{!}{\includegraphics{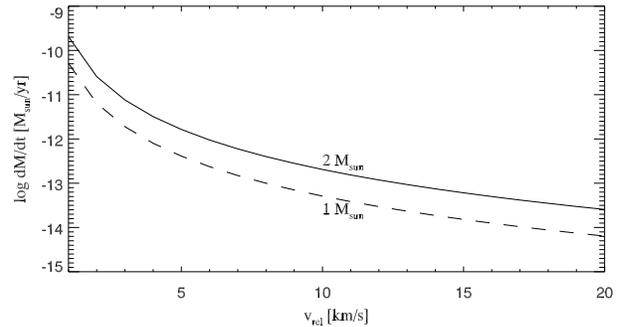}}
\caption{Mass accretion rates for different relative velocities between
ISM cloud and star. The results are shown for a star with 1~M$_\odot$
and a star with 2~M$_\odot$.}
\label{fig_mass_acc}
\end{figure}

\citet{Bondi} derive the radius for which ISM 
particles are accreted onto the star
\begin{equation}
r_{\rm acc} = \frac{\sqrt{2.5}\,GM}{v_{\rm rel}^2}~~~{\rm cm}\,\,\,.
\end{equation}
Different prefactors can be found in the literature, but \citet{Bondi} 
showed that $\sqrt{2.5}$ holds for steady state.
This radius leads to an accretion rate of
\begin{equation}
\dot{M} = \pi r_{\rm acc}^2 \rho v_{\rm rel}~~~{\rm g~s}^{-1}\,\,\,.
\end{equation}
An A-type star of 2~M$_\odot$ passing the diffuse ISM cloud at a relative 
velocity of 17~km~s$^{-1}$ has an accretion radius of 9.7~AU and an 
accretion rate of $4.2\times 10^{-14}$~M$_\odot$~yr$^{-1}$.
Since no evidence is found for emission lines in the ultraviolet 
spectral range and accretion must overcome diffusion, the accretion 
rate must lie between $10^{-9}$ \citep{Bertout} and 
$10^{-14}$~M$_\odot$~yr$^{-1}$ \citep{Turcotte}.
Hence the range of relative velocities between star and diffuse 
ISM cloud is 1 to 20~km~s$^{-1}$ (Fig.~\ref{fig_mass_acc}).

This estimate assumes that collisions occur frequently enough to
reduce the angular momentum of the particles streaming around the star.
In order to verify this, one can calculate the number of collisions
of a particle on a circular orbit with a radius $r_{\rm acc}$ around 
the star. We assume that collisions are effective in removing angular 
momentum over a path length of $\pi r_{\rm acc}$ and that the
cross section of the gas particle is $6.3\times 10^{-16}$~cm$^2$ 
(corresponding to a radius of $10^{-8}$~cm)
\begin{equation}
N_{\rm coll} = 6.3\times 10^{-16} n \frac{\sqrt{2.5}\,\pi GM}{v_{\rm rel}^2}\,\,\,,
\end{equation}
where $n$ denotes the density of the ISM cloud. Inserting a 2~M$_\odot$
star and assuming a relative velocity of 17~km~s$^{-1}$, we
obtain 3 collisions for the orbiting particle. 
The accretion scenario will certainly work for ISM clouds with 
densities $\ge 10$~cm$^{-3}$.

Typical sizes $d$ for ISM clouds are of the order of 0.1 to 10~pc
\citep{Dring}. The duration of accretion is determined by 
the relative velocity between cloud and star and the dimension of the cloud
\begin{equation}
t_{\rm acc} = 9.8\times 10^5 \left(\frac{d}{\rm pc}\right)
              \left(\frac{v_{\rm rel}}{\rm km/s}\right)^{-1}~~~{\rm yr}\,\,\,.
\end{equation}
In the above scenario with a relative velocity of 17~km~s$^{-1}$,
the star passing a diffuse ISM cloud of 1~pc accretes for 
$5.8\times 10^4$~yr. 

In the light of the above described scenario, the $\lambda$~Bootis
phenomenon is a transient phenomenon. The abundance pattern imprinted by 
the diffuse ISM cloud will disappear very rapidly, within $10^6$~yr, after 
the star has passed the cloud \citep{Turcotte}. 
This is due to diffusion taking over again and to effective mixing by 
convection and meridional circulation. Since the time, which a star 
actually spends crossing a diffuse ISM cloud, is very short, we do not 
expect circumstellar/interstellar lines towards every $\lambda$~Bootis 
star. In fact \citet{Holweger1} and \citet{Holweger2} have shown that 
about 30 per cent of the $\lambda$~Bootis stars reveal narrow absorption 
features in Ca\,{\sc ii}\,K. On the other hand \citet{Bohlender} have
found narrow Na\,{\sc i}\,D absorption features in 3 out of 8 
$\lambda$~Bootis stars.

\section{The borders of the $\lambda$~Bootis phenomenon}

The spectral range of $\lambda$~Bootis stars extends from B9.5\,V to 
F3\,V with a peak around F1\,V. 60 per cent of the $\lambda$~Bootis stars
are cooler than 8000~K and the stellar masses range between $1.56 
(8)$ and $2.50 (12)$~M$_\odot$ \citep{Paunzen02}. 
The distribution of rotational velocities of $\lambda$~Bootis A-type 
stars is indistinguishable from normal A-type stars 
\citep{Paunzen01}, the maximum so far being 250~km~s$^{-1}$.
We first consider the hot end of the phenomenon, where stellar winds 
stop the accretion of interstellar matter and then the 
cool end, where convection zones become too massive to be contaminated 
by the above derived accretion rates.

\subsection{The hot and cool end}
\label{hotcool}

B-type stars possess radiatively driven winds and typical values derived 
for the mass loss rates from 
radiative acceleration calculations by \citet{Abbott} can be found
in \citet{Cohen}. They range from 
$3\times 10^{-8}$~M$_\odot$~yr$^{-1}$ for a B0\,V star to 
$3\times 10^{-12}$~M$_\odot$~yr$^{-1}$ for a B7\,VIe star. 
\citet{Babel96} has shown that these calculations tend
to overestimate the mass loss rate by typically a factor of 4, because
they do not take into account the dependence of the radiative 
acceleration on the outwards velocity in the wind and the detailed
shadowing by photospheric lines. So the range of mass loss rates
for B-type stars will then become roughly $8\times 10^{-9}\ldots 
8\times 10^{-13}$~M$_\odot$~yr$^{-1}$.

For the A-type stars \citet{Babel95} has found mass loss rates below 
$10^{-16}$~M$_\odot$~yr$^{-1}$. This mass loss is due to selective winds 
that act only on the metals and the terminal velocity of this wind is 
rather high, about 6000~km~s$^{-1}$. Comparing the momenta of the wind 
and the accretion flow, we find that a typical accretion rate of 
$10^{-13}$~M$_\odot$~yr$^{-1}$ at velocities of 10~km~s$^{-1}$ wins 
over the selective wind. Such selective mass loss will therefore not 
affect the formation of a $\lambda$~Bootis abundance pattern.
Since we do not know exact mass loss rates for late B-type stars, we can
only conclude that the hot end of the $\lambda$~Bootis phenomenon
lies at the transition between A and B-type stars.

Stellar evolutionary models show that the mass of the convection
zone at $10^8$~yr (that is on the main-sequence) is $10^{-8}$, $10^{-10}$, 
and $10^{-11}$~M$_\odot$ for a 1.5, 1.7, and 2.5~M$_\odot$ star 
respectively \citep{Richard}. Overshoot is supposed
to enlarge this mass by less than a factor of 10
\citep{Freytag}. Given the accretion 
timescales and rates derived in Sect.~\ref{model}, the photospheres
of stars more massive than 1.5~M$_\odot$ can be contaminated by the
diffuse ISM cloud.

\subsection{The rotational velocity}

\citet{Charbonneau} have shown that meridional 
circulation does not completely wipe out the chemical abundance pattern 
established under the influence of diffusion in fast rotating stars. 
The timescale for mixing by meridional circulation is a fraction of the 
Eddington-Sweet time \citep{Turcotte}
\begin{equation}
t_{\rm m} = \frac{G^2 M_\ast^3}{L_\ast R_\ast^2 v_{\rm e}^2}~~~{\rm s}\,\,\,,
\end{equation}
where $v_{\rm e}$ denotes the equatorial rotational velocity.
For the A5 and F5 star we obtain a mixing time of $8.4\times 10^6$ and
$2.1\times 10^7$~yr for an equatorial velocity of 200~km~s$^{-1}$. 
This is much longer than we expect the accretion signature to
last after the accretion process has ceased.

\section{Statistics}

In order to test our proposed scenario with the observed number
of $\lambda$~Bootis type objects within a given space volume, we
have made the following analysis. \citet{Paunzen01} has estimated
that the current spectral classification resolution surveys in the
relevant spectral domain are complete up to 60\,pc. Within this
space volume we know of 8 well established $\lambda$~Bootis type
stars: HD\,30422, HD\,31295, HD\,74873, HD\,110411, HD\,125162, HD\,142703,
HD\,183324 and HD\,192640. 

Within a space volume of 60~pc, there are about 1100 ``normal''
type stars in the relevant spectral domain. Since most of the
spectral classification surveys are mainly devoted to single
objects, we discard all apparent spectroscopic binary
systems. Taking an average binary frequency of 30 per cent, we are
left with 330 apparent single objects.

As next step, we derive a rough estimate of the space
volume for which the densities of the ISM are high enough 
so that accretion of interstellar matter on to the stellar surface 
occurs. For the space volume considered here, we 
are confronted with the ``Local
Bubble'' which surrounds the Sun. It stretches out with a radius 
between 30 and 300~pc. Its characteristics, distribution and
kinematic are still not well understood \citep{Genova,Snowden}. 
From the maps of the Local Bubble
from \citet{Welsh} and \citet{Sfeir}, we
determine an average value of 3 per cent for which the densities of the
ISM are high enough. Notice that the published maps reflect the
characteristics on rather large scales only (about 10 to 20~pc).
The kinematically and chemically structure varies on scales which
are 10 times smaller than that \citep{Genova}.

So we are left with 10 stars within a space volume of 60~pc 
which can be accounted as $\lambda$ Bootis type objects.
This is very well in the range of the actual observed number
(8). However, this does, by no means, prove our scenario, but
naturally explains the low number of detected members.

Let us also make a remark about the (non-) detection of $\lambda$
Bootis type objects in open clusters \citep{Gray1,Gray2}.
Up to now only few members in the Orion OB1 association 
and NGC\,2264 (ages about 10$^{7}$~yr) are known although an extensive
survey by the group of R.O.~Gray in about two
dozen open clusters has been performed. We believe
that the lack of members in open clusters is due to a lack of gas to
accrete. \citet{Palla} have shown that the remnant gas from
the stellar formation dissipates in less than 10$^{7}$~yr. This
even holds for aggregates containing no massive stars. 

\section{Observational tests}

Some $\lambda$~Bootis stars are members of binary systems. In all cases, 
where a detailed abundance analysis has been done so far, both stars 
turned in fact out to be $\lambda$~Bootis stars 
\citep{Sturenburg,Iliev,Heiter}: 
HD\,84948\,A+B, HD\,171948\,A+B, HD\,193281/HD\,193256, HD\,198160/HD\,198161. 
This is naturally explained by the above described
scenario, because both stars pass through the same diffuse cloud and 
accrete interstellar gas. If both stars have a spectral type between 
late B and early F, they both appear afterwards as $\lambda$~Bootis 
stars. But, if one component has an earlier or later spectral type, it 
will not show the typical $\lambda$~Bootis abundance pattern. 

To test the above described scenario, it is crucial to 
analyze more $\lambda$~Bootis binary systems and to find systems consisting 
of an A-type star and a companion which is beyond the hot or cool
end as described in Sect.~\ref{hotcool}.

\section*{Acknowledgements}
We have greatly benefited from many helpful discussions with
Harm~Habing, Garrelt~Mellema, Hartmut~Holweger and Werner~Weiss. 
EP acknowledges partly support by the Fonds zur F\"{o}rderung der 
wissenschaftlichen Forschung, project P14984.

\bsp

\label{lastpage}


\begin{thebibliography}{99}
\bibitem[\protect\citeauthoryear{Abbott}{1982}]{Abbott}
Abbott D.C., 1982, ApJ, 259, 282
\bibitem[\protect\citeauthoryear{Artymowicz \& Clampin}{1997}]{Artymowicz}
Artymowicz P., Clampin M., 1997, ApJ, 490, 863  
\bibitem[\protect\citeauthoryear{Babel}{1995}]{Babel95}
Babel J., 1995, A\&A, 301, 823  
\bibitem[\protect\citeauthoryear{Babel}{1996}]{Babel96}
Babel J., 1996, A\&A, 309, 867  
\bibitem[\protect\citeauthoryear{Bertout}{1989}]{Bertout}
Bertout C., 1989, ARA\&A, 27, 351
\bibitem[\protect\citeauthoryear{Black}{1987}]{Black}
Black J.H., 1987, in Hollenbach D.J., Thronson H.A., eds., Interstellar
      Processes. Dordrecht, D. Reidel Publishing Company, p. 731
\bibitem[\protect\citeauthoryear{Bohlender et al.}{1999}]{Bohlender}
Bohlender D.A., Gonzalez J.-F., Matthews J.M., 1999, A\&A, 350, 553
\bibitem[\protect\citeauthoryear{Bondi \& Hoyle}{1944}]{Bondi}
Bondi H., Hoyle F., 1944, MNRAS, 104, 273  
\bibitem[\protect\citeauthoryear{Charbonneau \& Michaud}{1991}]{Charbonneau}
Charbonneau P., Michaud G., 1991, ApJ, 370, 693  
\bibitem[\protect\citeauthoryear{Cohen et al.}{1997}]{Cohen}
Cohen D.H., Cassinelli J.P., MacFarlane J.J. 1997, ApJ, 487, 867  
\bibitem[\protect\citeauthoryear{Cox}{2000}]{Cox}
Cox A.N., 2000, Allen's Astrophysical Quantities, Springer Verlag  
\bibitem[\protect\citeauthoryear{Draine \& Lee}{1984}]{Draine}
Draine B.T., Lee H.M., 1984, ApJ, 285, 89 
\bibitem[\protect\citeauthoryear{Dring et al.}{1996}]{Dring}
Dring A.R., Murthy J., Henry R.C., 1996, ApJ, 457, 764  
\bibitem[\protect\citeauthoryear{Freytag et al.}{1996 and Freytag private communication}]{Freytag}
Freytag B., Ludwig H.-G., Steffen M., 1996, A\&A, 313, 497
\bibitem[\protect\citeauthoryear{G{\'e}nova et al.}{1997}]{Genova}
G{\'e}nova R., Beckman J.E. Bowyer, S. Spicer T., 1997, ApJ, 484, 761
\bibitem[\protect\citeauthoryear{Gray \& Corbally}{1999}]{Gray1}
Gray R.O., Corbally C.J., 1999, AAS, 195, 4702 
\bibitem[\protect\citeauthoryear{Gray \& Corbally}{2002}]{Gray2}
Gray R.O., Corbally C.J., 2002, AJ, in press 
\bibitem[\protect\citeauthoryear{Heiter}{2002}]{Heiter}
Heiter U., 2002, A\&A, 381, 959
\bibitem[\protect\citeauthoryear{Holweger \& Rentzsch-Holm}{1995}]{Holweger1}
Holweger H., Rentzsch-Holm I., 1995, A\&A, 303, 819
\bibitem[\protect\citeauthoryear{Holweger et al.}{1999}]{Holweger2}
Holweger H., Hempel M., Kamp I., 1999, A\&A, 350, 603
\bibitem[\protect\citeauthoryear{Iliev et al.}{2002}]{Iliev}
Iliev I.Kh., Paunzen E., Barzova I.S. et al., 2002, A\&A, 381, 914  
\bibitem[\protect\citeauthoryear{Kurucz}{1992}]{Kurucz}
Kurucz R.L., 1992, Rev. Mex. Astron. Astrofis., 23, 181  
\bibitem[\protect\citeauthoryear{Palla \& Stahler}{2000}]{Palla}
Palla F., Stahler S.W., 2000, ApJ, 540, 255
\bibitem[\protect\citeauthoryear{Paunzen}{2001}]{Paunzen01}
Paunzen E., 2001, A\&A, 373, 633  
\bibitem[\protect\citeauthoryear{Paunzen et al.}{2002}]{Paunzen02}
Paunzen E., Iliev I.Kh., Kamp I., Barzova I.S., 2002, MNRAS, in press  
\bibitem[\protect\citeauthoryear{Richard et al.}{2001}]{Richard}
Richard O., Michaud G., Richer J., 2001, ApJ, 558, 377
\bibitem[\protect\citeauthoryear{Sfeir et al.}{1999}]{Sfeir}
Sfeir D.M., Lallement R., Crifo F., Welsh B.Y., 1999, A\&A, 346, 785
\bibitem[\protect\citeauthoryear{Snowden et al.}{1998}]{Snowden}
Snowden S.L., Egger R., Finkbeiner D.P., Freyberg M.J., Plucinsky P.P.,
 1998, ApJ, 493, 715
\bibitem[\protect\citeauthoryear{St{\"u}renburg}{1993}]{Sturenburg}
St\"{u}renburg S., 1993, A\&A, 277, 139  
\bibitem[\protect\citeauthoryear{Turcotte \& Charbonneau}{1993}]{Turcotte}
Turcotte S., Charbonneau P., 1993, ApJ, 413, 376   
\bibitem[\protect\citeauthoryear{Venn \& Lambert}{1990}]{Venn}
Venn K.A., Lambert D.L., 1990, ApJ, 363, 234
\bibitem[\protect\citeauthoryear{Vergely et al.}{2001}]{Vergely}
Vergely J.-L., Freire Ferrero R., Siebert A., Valette B., 2001, 
  A\&A, 366, 1016
\bibitem[\protect\citeauthoryear{Weingartner \& Draine}{2001}]{Weingartner}
Weingartner J.C., Draine B.T., 2001, ApJS, 134, 263
\bibitem[\protect\citeauthoryear{Welsh et al.}{1998}]{Welsh}
Welsh B.Y., Crifo F., Lallement R., 1998, A\&A, 333, 101
\end{thebibliography}
\end{document}